# 7

# Atomic Layer Deposition of High-k Oxides on Graphene

Harry Alles, Jaan Aarik, Jekaterina Kozlova,
Ahti Niilisk, Raul Rammula and Väino Sammelselg
*University of Tartu*
*Estonia*

## 1. Introduction

Graphene that is a single hexagonal layer of carbon atoms with very high intrinsic charge carrier mobility (more than 200 000 cm²/Vs at 4.2 K for suspended samples; Bolotin, et al., 2008) attracts attention as a promising material for future nanoelectronics. During last few years, significant advancement has been made in preparation of large-area graphene. The lateral sizes of substrates for graphene layers have been increased up to ¾ m (Bae et al., 2010) and continuous roll-to-roll deposition of graphene has been published (Hesjedal, 2011). This kind of progress might allow one to apply similar planar technologies for fabricating graphene-based devices in future as currently used for processing of silicon-based structures.

After very first experiments (Novoselov et al., 2004), in which the electrical properties of isolated graphene sheets were characterized, a lot of attention has been paid to the similar studies, i.e. investigation of uncovered graphene flakes deposited on oxidized silicon wafers that served as back gates. However, in order to realize graphene-based devices, a high-quality dielectric on top of graphene is required for electrostatic gates as well as for tunnel barriers for spin injection. For efficient control of charge carrier movement dielectric layers deposited on graphene should be very thin, a few nanometers thick, and of very uniform thickness without any pinholes. At the same time, the dielectric should possess high dielectric constant, high breakdown voltage and low leakage current even at a small thickness. And, of course, it is expected that the high mobility of charge carriers in graphene should not be markedly affected by the dielectric layer.

In order to make top-gated graphene-based Field Effect Transistor (FET), Lemme et al. (2007) applied evaporation techniques for preparation of a gate stack with ~20 nm thick $SiO_2$ dielectric layer on graphene. They used p-type Si(100) wafers with a boron doping concentration of $10^{15}$ cm$^{-3}$, which were oxydized to a $SiO_2$ thickness of 300 nm. On these wafers, micromechanically exfoliated graphene flakes were sticked. The Ti/Au source and drain electrodes were prepared using optical lift-off lithography. Next, electron beam lift-off lithography was applied to define a top gate electrode on top of the graphene flake covered with the dielectric (Fig. 1a).

Lemme et al. were first to demonstrate that the combined effect of back and top gates can be applied to graphene devices. However, measurements of the back-gate characteristics before and after the evaporation of the top gate revealed that deposition of the dielectric layer caused a considerable decrease in the drain current of the device (Fig. 1b). In addition, the current modulation through the back gate was reduced by the top gate. After this degradation the room-temperature electron and hole mobilities in graphene were as low as 710 cm$^2$/Vs for holes and 530 cm$^2$/Vs for electrons, i.e. comparable to the typical charge carrier mobilities in silicon.

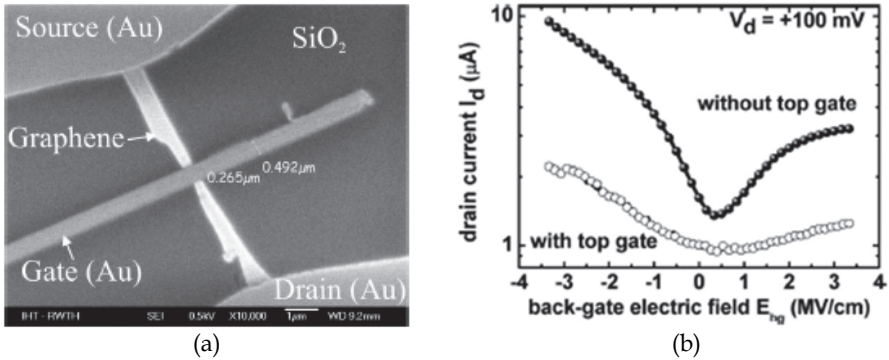

(a)  (b)

Fig. 1. (a) Scanning Electron Microscope image of a graphene FET. (b) Back-gate transfer characteristics of the graphene FET with and without a top gate (Adapted from Lemme et al., 2007).

Using Raman spectroscopy, Ni et al. (2008a) have studied the effect of deposition of different dielectrics with various methods on top of graphene sheets. Figure 2 presents Raman spectra after deposition of SiO$_2$ layer by Electron Beam Evaporation (EBE), Pulsed Laser Deposition (PLD) and Radio Frequency (RF) sputtering methods, HfO$_2$ layer by PLD and polymethyl methacrylate (PMMA) layer by spin coating on top of graphene. As it can be seen, the used deposition methods (with the exception of spin coating) caused defects which showed up as appearence of the D-band in Raman spectra (~ 1350 cm$^{-1}$) of graphene.

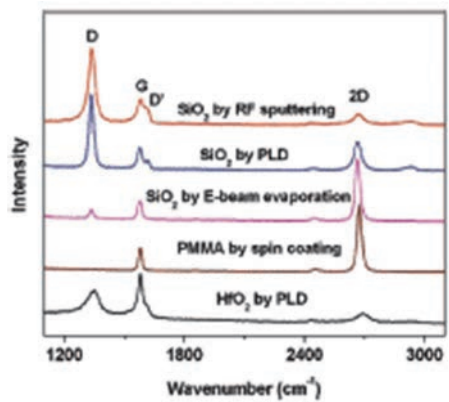

Fig. 2. Raman spectra of graphene after deposition of SiO$_2$ by RF sputtering, PLD and EBE, PMMA by spin coating and HfO$_2$ by PLD. (Adapted from Ni et al., 2008a).

On the basis of the Raman spectra, one can conclude that electron-beam evaporation and, particularly, sputtering and PLD processes may cause significant damages to graphene. Due to that influence the mobility values of the top-gated graphene-based devices prepared using these methods are typically by an order of magnitude smaller than those of back-gated devices (Lemme et al., 2007). For this reason, alternative methods for deposition of dielectrics on graphene have been extensively investigated.

In the case of traditional nanoelectronic devices, a very suitable method for controlled deposition of ultrathin homogeneous films is Atomic Layer Deposition (ALD). However, ALD of thin films on graphene is not easy because there are no dangling bonds on the defect-free graphene surface, which are needed for chemical surface reactions the conventional ALD processes are based on. Nevertheless, several groups have been able, using ALD technique, to deposit thin and continuous $HfO_2$ layers on pristine as-cleaved graphene (Meric et al., 2008; Zou et al., 2010; Alles et al., 2011). It has also been reported that after ALD of relatively thin $Al_2O_3$ layers directly on graphene (Moon et al., 2009; Nayfeh, 2011), top-gated devices with good performance can be obtained.

In order to better nucleate the growth of dielectric on graphene and obtain smooth uniform films, the graphene surface has been pretreated prior ALD (Williams et al., 2007), e.g. by using receipies successfully tested on carbon nanotubes (Farmer & Gordon, 2006). Later, other approaches for pretreatment of graphene surface have been applied as well. For instance, a thin (~1 nm thick) metal seed layer of Al that was oxydized before ALD (Kim et al., 2009) or polymer films (Wang et al., 2008; Farmer et al., 2009; Meric et al., 2011) have been deposited on graphene in order to initiate ALD of high-k dielectrics. In addition, pretreatment of the graphene surface with ozone prior ALD of $Al_2O_3$ has been investigated (Lee et al., 2008, 2010).

In this work, we compare ALD experiments in which graphene has been covered by high-k dielectrics ($Al_2O_3$ and $HfO_2$) either directly or after surface functionalization. Particular attention is focussed on attempts to grow dielectric films with higher density (and dielectric constant) using higher substrate temperatures and two-temperature ALD processes that start with formation of a thin seed layer at low temperature and proceed with depositing the rest of the dielectric layer at high temperature.

## 2. ALD of high-k dielectrics directly on graphene

### 2.1 ALD of amorphous $Al_2O_3$ and $HfO_2$ directly on graphene

Wang et al. (2008) have tried to deposit a thin $Al_2O_3$ layer, ~2 nm thick, on mechanically exfoliated graphene sheets, which were carefully cleaned by annealing at 600 °C in Ar atmosphere at a pressure of 1 Torr. The deposition of $Al_2O_3$ on graphene at 100 °C using vapors of trimethylaluminum (TMA; $Al(CH_3)_3$) and water ($H_2O$) as precursors was unsuccessful – the $Al_2O_3$ film was preferentially formed on graphene edges and defect sites. On the basis of these results, it was concluded that ALD of metal oxides gives no direct deposition on defect-free pristine graphene.

Similar results were obtained by Xuan et al. (2008) who tried to deposit $Al_2O_3$ and $HfO_2$ films on Highly Ordered Pyrolytic Graphite (HOPG) surfaces. Fresh HOPG surfaces, obtained using Scotch tape, were transferred into ALD reactor immediately after cleaving and 1-35 nm thick $Al_2O_3$ films were deposited at the temperature of 200–300 °C by alternating pulses of TMA and $H_2O$ as precursors. For ALD of $HfO_2$, $HfCl_4$ and $H_2O$ were used as precursors. As a result, Xuan et al. obtained large number of $Al_2O_3$ and $HfO_2$ nanoribbons, with dimensions of 5 - 200 nm in width and more than 50 μm in length. This was due to the existence of numerous step edges on HOPG surfaces, which served as nucleation centers for the ALD process.

In our ALD experiments, we deposited thin $Al_2O_3$ film directly onto exfoliated graphene using $AlCl_3$-$H_2O$ precursor combination. During the film growth, the precursor pulses and $N_2$ purge period after the $AlCl_3$ pulse were 2 seconds while the $N_2$ purge period after the $H_2O$ pulse was 5 seconds in duration. To initiate nucleation, 4 cycles were applied at a substrate temperature of 80 °C. Then the temperature was increased to 300 °C and 50 cycles were applied. It was found, however, that such a deposition process still provided non-uniform coverage with large number of pinholes (Fig. 3a). About 4 nm thick $Al_2O_3$ film on graphene had the rms surface roughness value of ~1.6 nm while the roughness value of the film on $SiO_2$ was only 0.45 nm. Atomic Force Microscope (AFM) measurements revealed that some of the pinholes penetrated through the whole film. These results clearly correspond to the previous reports (Wang et al., 2008; Xuan et al., 2008) and indicate that $Al_2O_3$ nucleation directly onto graphene is retarded due to the absence of favourable surface sites. At the same time the Raman spectra of the same sample, presented in Fig. 3b, reveal that the ALD process has not generated defects in graphene. However, blueshifts of G- and 2D-bands indicate that the deposition process and/or deposited film influenced properties of graphene. Possible reasons for this influence will be discussed in Sec. 2.2.

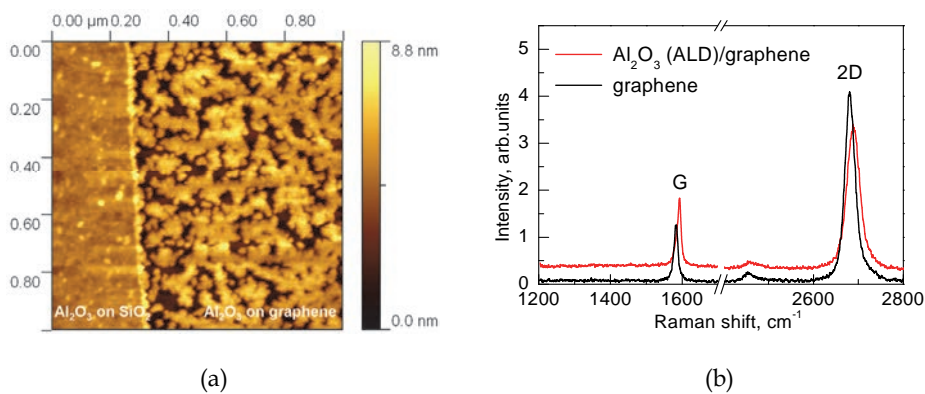

(a)          (b)

Fig. 3. (a) AFM image (1 × 1 μm²) of a single layer graphene flake (right) on $SiO_2$, both surfaces covered with ~4 nm thick $Al_2O_3$ layer and (b) typical Raman spectra from the same graphene flake before and after the ALD of $Al_2O_3$.

Very recently Nayfeh et al. (2011) reported about successful remote plasma-assisted ALD experiments in which $Al_2O_3$ gate dielectric was directly deposited onto chemical-vapor-deposited (CVD) monolayer graphene at 100 °C. They used 100 cycles consisting of TMA exposure for 0.06 s, purge for 30 s, oxygen (20 sccm) plasma exposure for 1 s and purge for 45 s. *In situ* annealing at 250 °C was also performed to improve $Al_2O_3$ film quality. AFM and Raman measurements showed a uniform dielectric coverage of graphene together with a slight increase in the disorder and signs of additional doping (see Fig. 4). This deposition process has enabled the fabrication of graphene FETs with ~9 nm thick gate insulator, which had a peak field-effect mobility of 720 cm²/Vs for holes.

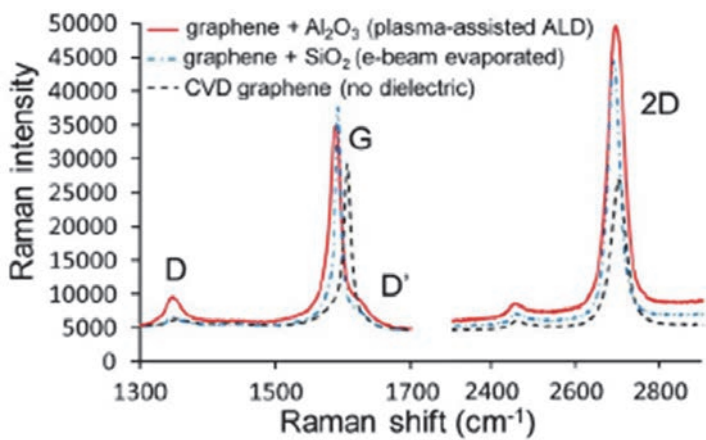

Fig. 4. Raman measurements of a CVD graphene before dielectric deposition and after 100 cycles of plasma-assisted ALD of $Al_2O_3$, and for a comparison, also after EBE of 9 nm $SiO_2$. (Adapted from Nayfeh et al., 2011).

A few years ago, Meric et al. (2008) reported about their experiments on ALD of 5-15 nm thick $HfO_2$ films on exfoliated graphene. They used tetrakis(dimethylamino)hafnium(IV) ($Hf(NMe_2)_4$) and water ($H_2O$) as the precursors and the deposition temperature was 90 °C. The pulse time for hafnium precursor was 0.3 s, which was followed by a 50 s purge, 0.03 s $H_2O$ pulse and 150 s purge. The resulting growth rate was about 0.1 nm/cycle and the dielectric constant of the $HfO_2$ film obtained was ~16 as determined by C-V measurements.

Figure 5a shows an AFM image of a single layer graphene flake on $SiO_2$ coated with a 5 nm thick $HfO_2$ layer grown immediately after mechanical exfoliation. As the measured height difference over the edge of the graphene is approximately the same (~0.9 nm) as before the ALD of $HfO_2$ layer, one can conclude that the growth has taken place at the same rate on the $SiO_2$ and graphene surfaces. Meric et al. suggested that the growth was most likely due to physiosorption, enhanced by the low-temperature growth procedure.

Meric et al. observed, however, that the roughness of the oxide on top of graphene was noticeably, by about 30%, higher than on the surrounding $SiO_2$. This result demonstrates that even at this very low temperature, uniformity of adsorption of precursors is not sufficient on graphene. Meric et al. also found that the mobility of the graphene sheet is almost the same before and after $HfO_2$ growth (Fig. 5b).

Recently, using the same precursors, pinhole-free 10 nm thick amorphous $HfO_2$ films have been deposited on exfoliated graphene at a temperature of 110 °C by Zou et al. (2010) who achieved the low-temperature mobility of charge carriers as high as ~20 000 $cm^2/Vs$ in $HfO_2$-covered graphene. To identify the growth mechanism, they have studied the deposition of $HfO_2$ films with various thicknesses on single- and multi-layer (5-6 layers) graphene flakes on $SiO_2$ substrates. On single-layer graphene flakes, they achieved the coverage of about 98% when depositing 2.5 nm thick $HfO_2$ films. Films thicker than 10 nm were pinhole-free and showed excellent morphology with the rms surface roughness of 0.2-0.3 nm, comparable to that of the $HfO_2$ film deposited on the $SiO_2$ substrate.

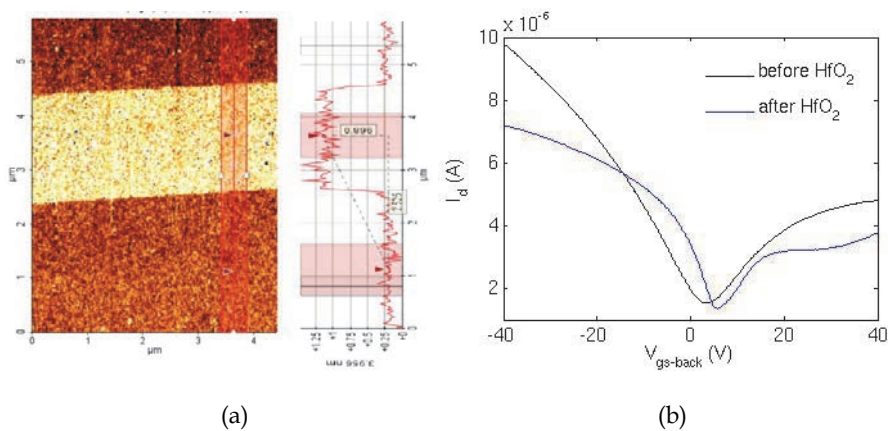

(a) (b)

Fig. 5. (a) AFM image and profile of a graphene flake covered with 5 nm $HfO_2$. (b) Back-gate transfer characteristics of the graphene-based FET (the gate width is 1.5 μm and the length is 3 μm) before and after the deposition of $HfO_2$. (Adapted from Meric et al., 2008).

On the other hand, $HfO_2$ films deposited on thicker graphene flakes exhibited much poorer quality as the coverage of 2.5 nm thick $HfO_2$ films on 5-6 layer graphene was only about 50% and pinholes remained even in 20 nm thick $HfO_2$ films. Zou et al. also found that the surfaces of single-layer graphene flakes were significantly rougher than those of multi-layer flakes. According to their explanation, the curvature existing in single-layer graphene flakes facilitates adsorption of the precursors. This explanation is in agreement with the observations of other groups who have recorded low coverage of the oxide layers on clean HOPG surfaces (Xuan et al., 2008).

### 2.2 ALD of crystalline $HfO_2$ films on as-cleaved graphene

The ALD processes described in the previous section have been performed at relatively low substrate temperatures. The films deposited at these temperatures are amorphous and for this reason possess relatively low dielectric constant. In order to obtain higher dielectric constant values, higher substrate temperatures yielding films with higher purity and density are required during the deposition of dielectrics. In our recent study (Alles et al., 2011), we have deposited $HfO_2$ films directly on as-cleaved graphene from $HfCl_4$ and $H_2O$ using three different processing regimes with substrate temperatures of (i) 180 °C for deposition of the whole dielectric layer, (ii) 170 °C for deposition of a 1 nm thick $HfO_2$ seed layer and 300 °C for deposition the rest of the dielectric layer, and (iii) 300 °C for deposition of the whole dielectric layer. An ALD cycle consisted of an $HfCl_4$ pulse (5 s in duration), purge of the reaction zone with $N_2$ (2 s), $H_2O$ pulse (2 s) and another purge (5 s).

Relatively smooth $HfO_2$ films were obtained on graphene at the low temperature of 180 °C. In the best cases, the rms surface roughness was less than 0.5 nm for ~30 nm thick films (see Fig. 6a). This value is comparable to the rms roughness of the same $HfO_2$ film on $SiO_2$, which was the substrate for the graphene flakes. The films were amorphous according to the Raman spectroscopy data (see Fig. 7).

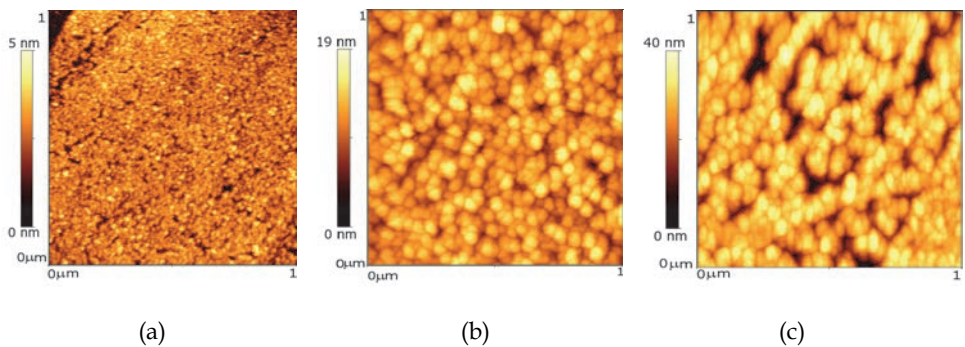

(a) (b) (c)

Fig. 6. AFM images of ~30 nm thick $HfO_2$ films on top of graphene flakes deposited at 180 °C (the rms surface roughness < 0.5 nm), (b) in a two-temperature (170/300 °C) process (the rms surface roughness ~2.5 nm) and (c) at 300 °C (the rms surface roughness ~5 nm).

The $HfO_2$ films deposited to a thickness of 30 nm in a two-temperature process had the rms surface roughness values of ~2.5 nm and ~2 nm on graphene and on $SiO_2$, respectively (Fig. 6b), and we recorded also Raman scattering from monoclinic $HfO_2$ (Fig. 7). At the same time, the surface roughness of ~30 nm thick $HfO_2$ films grown at 300 °C from the beginning of deposition process was as high as 5 nm on graphene and 2 nm on $SiO_2$. Scattering from monoclinic $HfO_2$ was recorded for these films as well (Fig. 7).

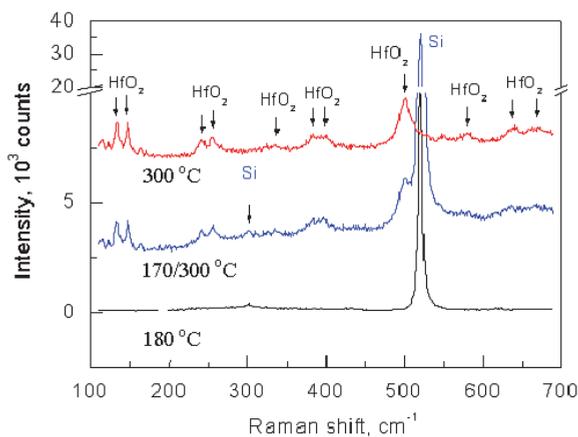

Fig. 7. Raman spectra from $HfO_2$ films on top of graphitic flakes deposited at 180 °C, 170/300 °C and 300 °C.

In our study, we found that the rms surface roughness of ~10 nm thick $HfO_2$ films grown on graphene was markedly lower (~1.7 nm) compared with that of ~30 nm thick $HfO_2$ films (~2.5 nm). Similar difference in surface roughness was observed for films deposited on $SiO_2$. Thus, the rougher surface of thicker $HfO_2$ films is due to crystallization of $HfO_2$ during the film growth at higher temperature rather than because of nucleation problems on the surface of graphene.

At the edge of a single layer graphene flake, the AFM surface profile of the $HfO_2$ film deposited in the two-temperature process (see Fig. 8) corresponded very well to the profile recorded before deposition of the dielectric. This means that $HfO_2$ film grew at the same rate on graphene and $SiO_2$. However, as the surface of $HfO_2$ films is still somewhat rougher on graphene compared to that on $SiO_2$, the parameters of ALD process and the thickness of the seed layer need further optimization.

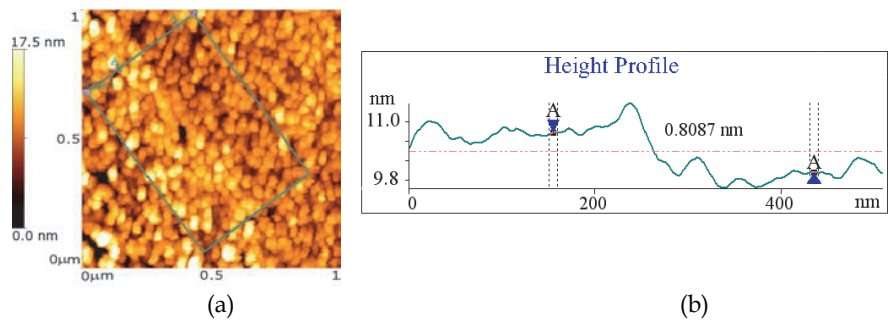

(a) (b)

Fig. 8. (a) AFM image of an edge of a single layer graphene flake after deposition of an 11 nm thick $HfO_2$ layer in a two-temperature process (the rms surface roughness is 1.7 nm on graphene and 0.9 nm on $SiO_2$). (b) AFM height profile over the edge shown in (a). (Adapted from Alles et al., 2011).

Figure 9 shows the Raman spectra of a single layer graphene flake taken at the same location before and after the deposition of an 11 nm thick $HfO_2$ layer in a two-temperature process. The intensities of peaks are lower after deposition and the background level has increased. But most importantly, the spectra reveal that noticeable blueshifts of G- (at ~1580 cm$^{-1}$) and 2D-bands (~2670 cm$^{-1}$) of graphene by 9 cm$^{-1}$ and 22 cm$^{-1}$, respectively, have appeared as a result of $HfO_2$ deposition.

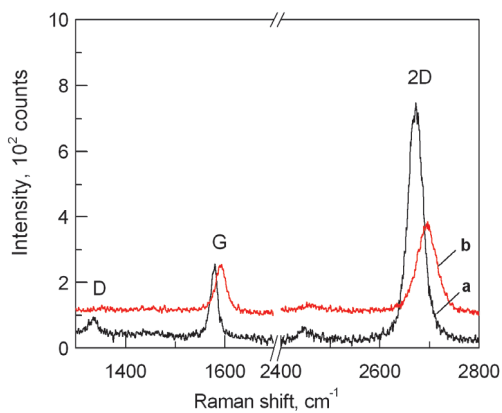

Fig. 9. Raman spectra of a single-layer graphene flake taken at the same point (a) before and (b) after the deposition of an 11 nm thick $HfO_2$ layer in a two-temperature (170/300 °C) process. (Adapted from Alles et al., 2011).

It is known that doping of graphene can influence the positions of Raman peaks (Das et al., 2008; Ni et al., 2008b) but the shift of the 2D-band in this case cannot be greater than that of G-band. In our experiments, however, the shift of the 2D-band is markedly greater. Edge effects and/or changes in the doping level can also cause changes in the peak positions but in that case narrowing of the G-peak should take place (Casiraghi et al., 2007; Ni et al., 2008b). In our case, on the contrary, the width of the G-peak increases. We also performed annealing experiments with a single layer graphene flake under conditions similar to those used during two-temperature ALD process of $HfO_2$. We observed blueshifts of G- and 2D-peaks by 4 cm$^{-1}$ and 7 cm$^{-1}$, respectively. At the same time full width at half maximum (FWHM) of the 2D-peak slightly increased by a few cm$^{-1}$, while FWHM of the G-peak decreased by ~4 cm$^{-1}$. On the basis of these data, doping of graphene, edge effects, and influence of high-temperature treatment during the ALD process could be excluded from the list of most important reasons for changes of Raman spectra caused by deposition of $HfO_2$.

Thus, the compressive strain developed in graphene during the two-temperature ALD process is the most probable reason for the blueshifts in the Raman spectra. Using the biaxial strain coefficient of –58 cm$^{-1}$/% for the Raman G-mode and –144 cm$^{-1}$/% for the Raman 2D-mode (Mohiuddin et al., 2009) and assuming elastic behavior of graphene, we estimated the compressive strain to be ~0.15% in our single layer graphene flake. This strain can be well explained by the relatively large and negative thermal expansion coefficient of graphene (7 x 10$^{-6}$ K$^{-1}$ at room temperature; Bao et al., 2009) while the thermal expansion of $HfO_2$ is of the same magnitude but with opposite (positive) sign (Wang et al., 1992), and it indicates strong adhesion of $HfO_2$ to graphene.

## 3. ALD of high-k dielectrics on functionalized graphene

### 3.1 ALD of $Al_2O_3$ on graphene after treatment with $NO_2$ and TMA

In order to reduce the leakage currents through the $Al_2O_3$ gate dielectric deposited on graphene, Williams et al. (2007) adopted the method proposed by Farmer & Gordon (2006) for pretreatment of carbon nanotubes and used $NO_2$ and TMA for functionalization of the graphene surface prior ALD of $Al_2O_3$. The exfoliated graphene flakes were cleaned with acetone and isopropyl alcohol (IPA) immediately before inserting them into the ALD reactor. Next, after the chamber was pumped down to a pressure of 0.3 Torr, non-covalent functionalization layer (NCFL) was deposited at room temperature using 50 cycles of $NO_2$ and TMA followed by 5 cycles of $H_2O$-TMA in order to prevent desorption of the NCFL. Finally, $Al_2O_3$ was grown at 225 °C with 300 ALD cycles, each of those consisting of a pulse of $H_2O$ vapor (1 Torr, 0.2 s) and a pulse of TMA vapor (1.5 Torr, 0.1 s), under continuous flow of $N_2$ and with 5 s intervals between pulses. As a result, ~30 nm thick oxide layer was obtained on top of graphene consisting of NCFL and $Al_2O_3$ and having a mean dielectric constant $k$ ~ 6.

The same receipe was later also used by Lin et al. (2009), who fabricated top-gated graphene-based FETs operating at gigahertz frequencies (up to 26 GHz). They functionalized the surface of exfoliated graphene with 50 cycles of $NO_2$-TMA and deposited after that a 12 nm thick $Al_2O_3$ layer as the gate insulator. The dielectric constant of the oxide layers was determined by C-V measurements to be ~7.5. However, in their experiments severe degradation of measured mobilities (down to ~400 cm$^2$/Vs) was observed. Consequently, although this kind of functionalization process allows one to deposit thin (of the order of 10 nm) pinhole-free gate dielectrics on graphene by ALD, this can cause a significant reduction of field-effect mobility of charge carriers and, correspondingly, the channel conductance of graphene-based FETs.

### 3.2 ALD of $Al_2O_3$ on graphene after ozone treatment

Lee et al. (2008, 2010) investigated coating of exfoliated graphene and HOPG surfaces with $Al_2O_3$ layer in $O_3$-based ALD process. First, they used freshly cleaved HOPG surfaces and found that while TMA-$H_2O$ process caused selective deposition of $Al_2O_3$ only along step edges (as in the experiments of Xuan et al. (2008)), the TMA/$O_3$ process began to provide nucleation sites on basal planes of HOPG surface. Lee et al. (2008) proposed that the chemically inert HOPG surface was converted to hydrophilic mainly through epoxide functionalization.

Later, in order to deposit a uniform $Al_2O_3$ dielectric layer on top of a graphene flake, they used an ALD seed layer (~1 nm in thickness) grown by applying 6 cycles of TMA and $O_3$ at 25 °C. After that 155 cycles of TMA/$H_2O$ at 200 °C were applied to deposit additional $Al_2O_3$ layer (~15 nm in thickness). For their films, they obtained the dielectric constant $k \sim 8$, which is higher than the value typically reported for the $Al_2O_3$ films deposited on graphene flakes. The mobilities of their devices reached ~5000 cm²/Vs. Lee et al. (2010) also found out that an $O_3$ treatment at 25 °C introduces minor amount of defects in a single layer graphene, while a substantial number of defects appear at 200 °C (Fig. 10).

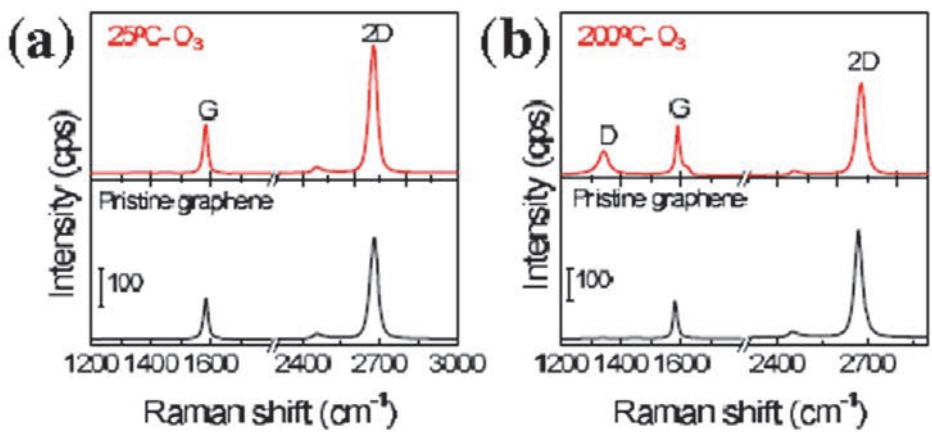

Fig. 10. Raman spectra of a pristine single layer graphene (bottom) and after treatment (top) (a) with $O_3$ at 25 °C and (b) with $O_3$ at 200 °C for 20 s. (Adapted from Lee et al., 2010).

### 3.3 Metal seed layer for ALD of $Al_2O_3$ on graphene

In order to make graphene-based top-gated FETs, Kim et al. (2009) first annealed the chips with exfoliated graphene flakes in a hydrogen atmosphere at 200 °C and then deposited a thin layer of Al by e-beam evaporation. After oxidization this layer served as a nucleation layer to enable ALD of $Al_2O_3$. TMA and $H_2O$ were used as precursors and 167 cycles resulted in about 15-nm thick $Al_2O_3$ layer on graphene. The devices fabricated using this technique indicated the mobility in excess of 6000 $cm^2$/Vs at room temperature. Consequently, the top-gate stack did not increase the carrier scattering significantly. A similar approach has later been used by several other groups, also with epitaxial graphene (Robinson, et al., 2010) and CVD graphene (Wu et al., 2011).

### 3.4 Polymer buffer layer for ALD of $Al_2O_3$ on graphene

Wang et al. (2008) used a polymer film as a buffer layer in order to cover carefully cleaned graphene with a very thin (~2-3 nm) $Al_2O_3$ layer. They soaked the chip with graphene flakes in 3,4,9,10-perylene tetracarboxylic acid (PTCA) solution for ~30 min, rinsed thoroughly and blew dry. The chip was then immediately moved into the ALD reactor and $Al_2O_3$ was deposited from TMA and $H_2O$ at 100 °C. Figure 11 shows the AFM images of a graphene flake before and after ALD. Uniform coverage with an ultrathin $Al_2O_3$ film was achieved as the measured rms surface roughness of $Al_2O_3$ on graphene was as low as 0.33 nm.

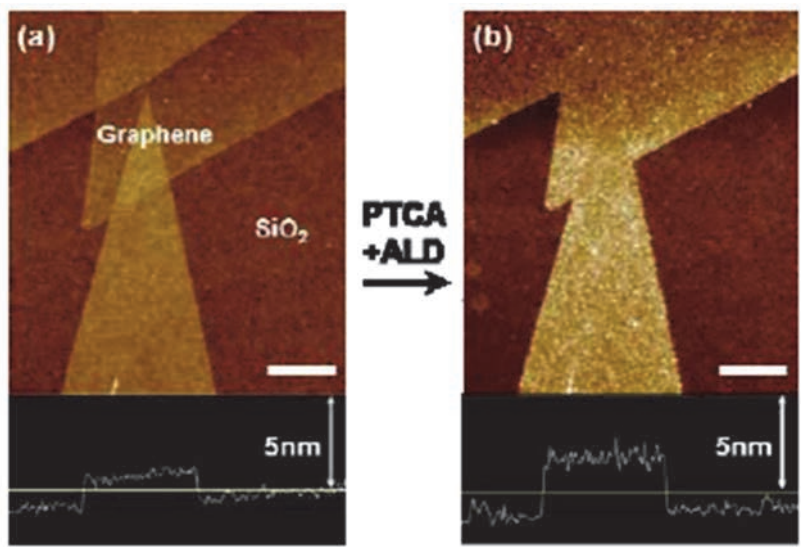

Fig. 11. AFM images of graphene (a) before ALD and (b) of the same area after ~2 nm $Al_2O_3$ deposition. Scale bar is 500 nm. (Adapted from Wang et al., 2008).

### 3.5 Metal seed layers for ALD of $HfO_2$ on graphene

Fallahazad et al. (2010) have investigated the carrier mobility in a single layer and bilayer exfoliated graphene with a top $HfO_2$ dielectric as a function of the $HfO_2$ film thickness and temperature. Prior to the $HfO_2$ film with ALD technique, a thin (~1.5 nm) seed layer of Al was deposited by e-beam evaporation. The $HfO_2$ layer was deposited at 200 °C from Hf(NMe$_2$)$_4$ and H$_2$O as presursors, without any postdeposition annealing. The relative dielectric constant of the stack was found to be about 16. A considerable mobility reduction to about 50% of the initial value was observed after the first 2-4 nm of metal oxide deposition. The mobility did not depend significantly on temperature in the range from 77 K to room temperature. This result suggests that phonon scattering did not play an essential role in the devices. Therefore the authors of this study speculated that influence of positively charged oxygen vacancies, ubiquitous in high-k dielectrics, was the main mobility limiting factor.

### 3.6 Polymer buffer layer for ALD of HfO$_2$ on graphene

Farmer et al. (2009) used a low-k polymer (NFC 1400-3CP) as a buffer layer for ALD of HfO$_2$ on exfoliated graphene. This polymer is a derivative of polyhydroxystyrene that is commonly used in lithography. The polymer can be diluted in propylene glycol monomethyl ether acetate (PGMEA) and spin-coated on top of graphene. For deposition of HfO$_2$, Farmer et al. used Hf(NMe$_2$)$_4$ and H$_2$O as precursors and carried the ALD process out at 125 °C. The process yielded HfO$_2$ films with a dielectric constant of about 13 on graphene.

Farmer et al. also found that in order to produce continuous functionalization layer on graphene, a 24:1 dilution (by volume) of PGMEA/NFC is sufficient. Spinning of such a solution at a rate of 4000 rpm for 60 s results in a layer of about 10 nm in thickness. After curing the buffer layer at 175 °C for 5 min to remove residual solvent, a 10-nm thick HfO$_2$ layer was deposited on top of that. The dielectric constant of the buffer layer was determined to be 2.4 which is a resonable value for this polymer.

The receipe of Farmer et al. was also used by Lin et al. (2010) who demonstrated cutoff-frequency as high as 100 GHz for top-gated FETs based on wafer-scale epitaxial graphene made from SiC. This cut-off frequency exceeds that of Si metal-oxide semiconductor FETs of the same gate length (~40 GHz at 240 nm). The carrier mobilities were maintained between 900 to 1520 cm$^2$/Vs across the 2-inch wafer. These values were largely the same as before the deposition of a top gate stack.

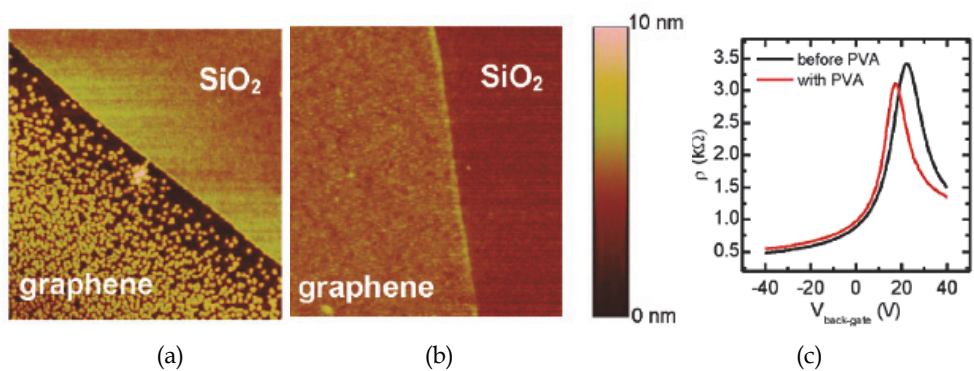

Fig. 12. (a) AFM image over an edge of a graphene flake on SiO$_2$ after ALD of a 5 nm thick HfO$_2$ (a) without and (b) with PVA. (c) Resistivity of a device after annealing and the PVA deposition as a function of back-gate bias (Adapted from Meric et al., 2011).

In their later experiments, Meric et al. (2011) have used PVA (poly(vinyl)alcohol) to provide a surface to seed ALD growth on exfoliated graphene. PVA is known to have a relatively high dielectric constant ($k \sim 6$). After the contact deposition they first cleaned the graphene surface by annealing at 330 °C for 3 h in forming gas and then dipped into an 1% aqueous solution of PVA for 12 h. This procedure resulted in ~2.5 nm thick layer of PVA on the graphene surface. After that a brief UV/ozone treatment was employed to activate the –OH groups before the ALD growth of $HfO_2$ at 150 °C with $Hf(NMe_2)_4$ and $H_2O$ for 50 cycles, yielding a 5-nm thick $HfO_2$ film (see Fig. 12b). As a result, the doping and the mobility of the top-gated graphene-based FET stayed relatively unchanged (see Fig. 12c).

Meric et al. (2011) also found that without PVA, ALD growth proceeds only on the $SiO_2$ area (see Fig. 12a) and the occasional patches of oxide on graphene are formed most likely by surface contamination.

## 4. Conclusion

As integration of graphene with high-quality ultrathin dielectrics is very important in development of graphene-based nanoelectronic devices, significant efforts have been concentrated on ALD of dielectric films on the graphene surface. An expected result of these studies is that at substrate temperatures most frequently used for ALD, i.e. at 200–400 °C, the deposition of uniform dielectric layers on clean surface of graphene is not possible due to the chemical inertness of this surface. However, using low-temperature ALD, several research groups have succeeded to grow dielectrics (e.g. $Al_2O_3$ and $HfO_2$) even on this kind of surfaces. Unfortunately, the quality of these films is usually not very high and/or the deposition process has a significant negative effect on properties of graphene. Thus, in order to cover graphene with uniform high-quality dielectric layers, different approaches to initiate the film growth have been investigated. It has been demonstrated that a seed layer can be grown by ALD on graphene using highly reactive precursors and very low deposition temperatures close to room temperature. Another way is to functionalize the graphene surface by deposition of metal or polymer buffer layers. It has to be noted, however, that functionalization, for instance with a metal seed layer can lead to degradation of the electronic properties of graphene. On the other hand, in the case of polymeric buffer layers, even when the electronic properties of graphene are not affected much, the total thickness of polymer/high-k dielectric layer might be too big for some kind of applications. Thus, the processes for deposition of dielectrics on graphene definitely need further optimization. It should also be pointed out that only a limited number of ALD experiments have been performed on CVD graphene, which has a great potential as a material for future nanoelectronics.

## 5. Acknowledgment

We would like to thank Pertti Hakonen for supporting the initiation of graphene studies at the Institute of Physics of the University of Tartu, Kaupo Kukli for useful discussions and Aleks Aidla and Alma-Asta Kiisler for technical assistance in experiments. This work was supported by Estonian Science Foundation (Grants No. 6651, 6999, 7845 and 8666), Estonian Ministry of Education and Research (targeted project SF0180046s07) and European Social Fund (Grant MTT1).